\documentclass{aa}
\usepackage{graphicx}
\usepackage{natbib}
\usepackage{amsmath}
\usepackage{amsfonts}
\bibpunct{(}{)}{;}{a}{}{,}

\newcommand{\rupa}{{\it VA~}}
\newcommand{\rupat}{{\it VA}}
\newcommand{\astroph}{\texttt{astro-ph}}

\begin{document}

  \title{Some comments on the note \\ 
    ``{\it Some comments on the paper \\ 
      ``Filter design for the detection of compact
      sources based on the Neyman-Pearson detector'' \\ 
      by
      M. L\'opez-Caniego et. al (2005, MNRAS 359, 993)}'' \\
    by R. Vio and P. Andreani \\
    (\astroph/0509394)}
  
   \author{M. L\'opez-Caniego\inst{1,2}
	  \and
	  D. Herranz\inst{1}
          \and
          R.B. Barreiro\inst{1}
	  \and
	  J.L. Sanz\inst{1}
          }

   \offprints{M. L\'opez-Caniego}

   \institute{Instituto de F\'\i sica de Cantabria,
     CSIC-UC, 
     Av. los Castros s/n,
     39005 Santander, Spain  \\
     \email{caniego@ifca.unican.es}
     \and
     Departamento de F\'\i sica Moderna,
     Universidad de Cantabria,
     Facultad de Ciencias,      
     Av. los Castros s/n, 
     39005 Santander, Spain}    

   \date{Received .............; accepted ................}
   
   \abstract{In this note we stress the necessity of a careful check
     of the arguments used by \citet{viodet} (\rupa hereinafter) to
     criticise the superior performance of the {\it biparametric scale
     adaptive filter} (BSAF) with respect to the classic {\it matched
     filter} (MF) in the detection of sources on a random Gaussian
     background. In particular, we point out that a defective reading
     and understanding of previous works in the literature
     \citep{ric,bar03,lop05} leads the authors of \rupa to the
     derivation of an incorrect formula and to some misleading
     conclusions.  \keywords{Methods: data analysis -- Methods:
     statistical} }
     
\titlerunning{Optimal Detection of Sources}
\authorrunning{M. L\'opez-Caniego et al.}
\maketitle

\section{Introduction}

In recent times, some controversy has arisen about the design of
\emph{``optimal''} filters for the detection of sources embedded in a
noisy background. The controversy seems to be focused on the following
question: do we have already an optimal tool for detecting such sources
or is it worth trying to find better methods for the task?  

In \citet{lop05} and some previous works \citep{bar03,lop04,lop05b} we
have explored the detection problem in the context of astronomical
data mining. The motivation of our work has been the need to detect
extragalactic objects, often referred to as \emph{``point sources''}
due to their small angular size, in microwave Astronomy. Since the
number of these objects increases very quickly as their flux
decreases, even a small improvement in our ability to notice faint
extragalactic objects can lead to a significant rise in the number of
detections.  Hence the importance of working out new and more powerful
detection procedures.

\citet{lop05} have proposed a detection procedure based on a common
practice in Astronomy, that consists in identifying possible sources
through the presence of ``peaks'' in the data. Commonly, the data is
previously filtered in order to improve the detectability of the
sources. Then, some decision rule is applied on the peaks in order to
determine whether they correspond to sources or not. 

A typical decision rule is based on the idea of amplitude
\emph{thresholding}, that is, the hypothesis that a source is present
in any considered point is accepted if the amplitude of the observed
data at that point is higher than a certain value. A decision rule
based only on the amplitude at the point where the decision has to be
made is missing information on the local structure of the source and
the background where it is embedded. Thus, in order to increase the
power of the decision rule, in \citet{lop05} we considered not only
the amplitude of the peaks, but also the curvature.

Following this idea, in \citet{lop05} we considered as decision rule
the Neyman-Pearson detector, which gives the highest number of
detections for a given number of false alarms; since the
Neyman-Pearson detector was to be applied to local maxima (peaks),
first we derived the expressions for the number density of peaks in an
interval $(x+dx)$ with amplitude $(\nu + d\nu)$ and curvature $(\kappa
+ d \kappa)$ both in presence and in absence of a source. These number
densities depend on the properties of the background, and these
properties can be modified, up to a certain extent, through filtering.
Then, we explored the performance of a linear filter, the Biparametric
Scale-Adaptive Filter (BSAF), that depends on a small number of
parameters that can be chosen so that the performance of the
Neyman-Pearson detector is optimised. We found that the BSAF
outperforms other filters such as the Mexican Hat Wavelet and the
standard matched filter (MF) in some interesting cases.

Very recently, \citet{viodet} have criticised the work presented in
\citet{lop05}.  We feel compelled to warn the community against some
of their remarks, that result to be either fruit of a bad
interpretation of our work or just plainly wrong.

\section{Some comments on the comments by \rupa}

In the introduction of their note, \rupa reproduce some very
well-known results about the matched filter in the context of the
Neyman-Pearson theorem, which can be found in any basic signal
detection textbook (for example \citealt{wa62}). Though this is
unquestionably correct, only the amplitude of the signal is considered
and, besides, it is still necessary to provide a criterion to localise
and define a single source among the set of pixels above the
threshold. We have followed a different approach that incorporates
the identification of any single source through the presence of
a local maximum and information about the curvature. Hence, our work
is not in contradiction with the scheme reproduced by \rupat, because
we are following a different, more complete, approach.

In addition, \rupa make three comments about our approach. In the
first comment, \rupa point out that one of our equations (equation (8)
in \citet{lop05}) is not correct. However, this statement is not true
and the alternative equation they propose is actually wrong.

In our work, we construct the Neyman-Pearson detector using the number
density of maxima of the background and the same number in the
presence of background and source. \citet{ric} obtains the number
density of maxima in an interval $(x+dx)$ with amplitude $(\nu+d\nu)$
and curvature $(\kappa + d \kappa)$ for a Gaussian background as:
\begin{equation}
n_b(\nu ,\kappa )  = \frac{n_b}{\sqrt{2\pi} }\frac{\kappa}{\sqrt{1-\rho^2}} 
\exp \left[
- \frac{\nu^2 + \kappa^2 - 2\rho \nu \kappa}{2(1 - \rho^2)} \right]
, 
\label{nbackground}
\end{equation}
where $\nu \in (-\infty,+\infty)$ and $\kappa \in (0,+\infty)$. Note
that the probability density $p_b(\nu,\kappa)$ is straightforwardly
obtained by dividing the previous equation by the total number density
$n_b$.

To obtain the probability density in the presence of a
point source of amplitude $\nu_s$ and curvature $\kappa_s$, \rupa simply
substitute $\nu \to \nu - \nu_s$ and  $\kappa \to \kappa - \kappa_s$
in $p_b(\nu,\kappa)$: 
\begin{align}  
& p(\nu ,\kappa |\nu_s)  =
\frac{1}{\sqrt{2\pi}}\frac{\kappa-\kappa_s}{\sqrt{1 - 
\rho^2}} \nonumber \\
& \times \exp \left[ - \frac{(\nu - \nu_s)^2 + (\kappa -
\kappa_s)^2 -  2\rho (\nu - \nu_s)(\kappa - \kappa_s)}{2(1 - \rho^2)} \right],
\label{nsource}
\end{align}
and indicate that $\nu \in (-\infty,+\infty)$ and $\kappa \in
(\kappa_s,+\infty)$. However, the derivation of this equation can not
be done in such a simple way. First of all, one needs to construct the
joint probability density of the field, its first and its second
derivative (where terms of the form $\nu-\nu_s$ and $\kappa -
\kappa_s$ appear). From this joint probability, one follows the
procedure explained in \citet{ric}, \citet{bar86} and \citet{bond87}
obtaining the number density of maxima in the intervals $(x+dx)$,
$(\nu + d\nu)$ and $(\kappa +d \kappa)$:
\begin{align}
& n(\nu ,\kappa |\nu_s) = \frac{n_b}{\sqrt{2\pi}}\frac{\kappa}{\sqrt{1
- \rho^2}} \nonumber \\ 
& \times \exp \left[- \frac{(\nu - \nu_s)^2 +
(\kappa - \kappa_s)^2 - 2\rho (\nu - \nu_s)(\kappa - \kappa_s)}{2(1 -
\rho^2)} \right],
\label{psource}
\end{align}
where $\nu \in (-\infty,+\infty)$ and $\kappa \in (0,+\infty)$.  Note
that the factor $\kappa$ that multiplies the exponential comes in from
imposing the condition of having a maximum and it refers to the total
curvature given by the background plus source (not only to the
background as stated by \rupat). We would like to remark that equation
(\ref{psource}) gives the number density of maxima coming from the
combination of the background plus source. This does not mean, at all,
that the maximum of the source has to coincide with a maximum of the
noise process as stated by \rupat. In addition, \rupa claims that
$\kappa \in (\kappa_s,+\infty)$, since they wrongly assume that the
maximum of the global field has to coincide with a maximum of the
background. However this is not true and therefore there is no reason
to restrict $\kappa$ to such interval. In fact $\kappa$ can take
values from $(0,+\infty)$. Note that this is another indication of the
fact that equation (\ref{nsource}) proposed by \rupa is wrong, since
this probability can take negative values when considering the correct
interval for $\kappa$.

Regarding the second comment of \rupat, they criticise the fact that we
work on a filtered version of the original signal. We would like to
stress that the common procedure in astronomy (and other fields) for
object detection is to filter the original image in order to enhance
the sources and then detect and identify those sources. Thus, an
important issue is not only to find the optimal filter, but also
which is the criterion to identify the sources. In our approach, we a
priori identify the maxima of the filtered image as source
candidates. Then, we apply a Neyman-Pearson detector to decide whether
the maximum is due or not to the presence of a point source. Note
that since we are considering the identification through the idea of
maxima, it is natural to define the Neyman-Pearson detector in terms
of number density of maxima. Taking into account these ideas, we
explore different filters and find that the BSAF outperforms the other
filters (including the MF) in some cases. 
Furthermore, \rupa suggest that we are trying to find an
approximate solution to the decision problem based on the likelihood
ratio
\begin{equation}
L(\mathbf{x},\nu,\kappa)=
\frac{p(\mathbf{x},\nu,\kappa|H_1)}{p(\mathbf{x},\nu,\kappa|H_0)}  > \gamma
\end{equation}
It is not clear to us what \rupa mean with this notation. If
${\mathbf{x}}$ is the observed 1-dimensional signal (as \rupa defined
in their introduction), $\nu$ is redundant. Also, if $\kappa$ refers
to the whole image, it should be a vector, $\boldsymbol{\kappa}$. We
understand that what \rupa mean is to construct a likelihood ratio
using the amplitude and curvature of all the pixels in the image (in
fact, if using all the pixels, one should also introduce the
information on the first derivative). However, this procedure does not
make sense in our approach, since we are considering only the maxima
of the image. Note that in the approach suggested by \rupat, one would
also need, a posteriori, a criterion to identify which points of the
image (from those above the threshold $\gamma$) correspond to each
source.  In addition, \rupa claim that our conclusions are drawn only
on the basis of numerical experiments. However, most of \citet{lop05}
is devoted to present the theoretical framework of our
method. Simulations are then performed in order to test the
theoretical results.  Therefore, the criticisms of \rupa in their
second comment are not well founded.

The third comment of \rupa refers to the procedure followed in the
numerical experiments. They criticise that we consider only those
sources whose peak is not moved to another pixel. We would like to
remark that the aim of our work was to present a novel theoretical
framework for object detection and to test it with numerical
simulations. Therefore, we try to reproduce exactly the theoretical
scheme with our simulations and focus only on what happens in one
pixel of the image, the pixel in which the source is located. In fact,
this point was already discussed in \citet{bar03}, finding that the
different filters there considered lead to similar number of
detections in the neighbouring pixels of the source and, thus, it did
not affect the conclusions.  In any case, in the more realistic case
when all the pixels of the image are considered, the conclusion that
the BSAF detects more sources than the other filters in the correct
localisation remains true.

Finally, \rupa comment in their conclusions that the performance of our
filter is based on strong a priori assumptions such as the Gaussianity
of the background and the symmetry of the source profile. We would
like to remark that many real fields do follow a Gaussian distribution
and therefore this is a very common and realistic assumption. In fact,
in their introduction, \rupa also assume the Gaussianity of the
background to show that the statistic given by the Neyman-Pearson
detector (when only information about the amplitude is used) leads to the
MF. Regarding the symmetry of the source profile, the filters can be
generalised without any difficulty to non-symmetrical profiles.

\section{Conclusions}

In a note recently appeared in \astroph, \rupa have made some comments
about our work \citep{lop05}. In this note, we have carefully checked
their arguments. The main comments made by \rupa are three. Let us
summarise:

In their first comment, \rupa have questioned an allegedly unproven
formula in our work, which is in fact rigorously derived from previous
works in the literature \citep{ric,bar86,bond87}. Instead, \rupa have
proposed an incorrect formula.

In their second comment, \rupa criticise the lack of generality of the
approach proposed in \citet{lop05}. In particular, \rupa criticise the
idea of filtering the data and applying the Neyman-Pearson detector to
the local maxima. Instead, they suggest that a generalisation of the
derivation of the Neyman-Pearson detector, including not only
amplitudes but also the second derivatives of the field, should be
done on a purely theoretical basis. Nevertheless, they are not able to
provide such a theoretical derivation, and the likelihood ratio they
propose is not general either, since it does not include the first
derivative of the field, that outside the maxima is not zero.  Our
approach, however, is consistent and it leads to an improvement in the
number of detections.

In their third comment, \rupa criticise a set of numerical experiments
designed to test our theoretical arguments precisely for being
designed to test our theoretical arguments. They suggest instead to
make numerical experiments in order to test what the theory does not
say. We have derived the number densities of maxima in two cases: when
a source is located at the position of the maxima (not ``nearby the
maxima'') and when there is no source. The way to test the hypothesis
expressed by these formulae is exactly the one explained in
\citet{lop05}. 

Besides the three main comments mentioned above, \rupa made a few
others.  One of them is that \rupa claim that our conclusions are
drawn only on the basis of numerical experiments, which is plainly
false. In \citet{lop05} we give a theoretical foundation for our
method, we make predictions based on the theory and then we check
those predictions with numerical simulations. The agreement is
excellent.

Other main objection is that our proposed method seems rather
complicated. Though it is true that simplicity is an aesthetically
admirable quality, we feel that a little complexity should not scare
scientists in their work. As mentioned in the introduction of this
note, it is worth to work hard to improve the capability of detection
of our statistical methods, even if the improvement is a few percent,
because it may lead to a significant rise in the number of
extragalactic objects detected.

Finally, \rupa blame our method for doing stringent a priori
assumptions, namely two: symmetry of the source profile and
Gaussianity of the background. It is false that our method requires
symmetry of the source: it was assumed only for simplicity but the
filters can be generalised to non-symmetric profiles just as the
standard matched filter can. Regarding Gaussianity, \rupa make in
their introduction exactly the same assumption as we do.

\end{document}